\documentclass[aps,prb, reprint,twocolumn,showpacs,superscriptaddress,preprintnumbers,amsmath,amssymb]{revtex4-1}

\usepackage{graphicx}
\usepackage{dcolumn}
\usepackage{bm}
\usepackage{textcomp}
\usepackage{upgreek}
\usepackage[colorlinks,pdfpagelabels,pdfstartview = FitH,bookmarksopen = true,bookmarksnumbered = true,linkcolor = black,plainpages = false,hypertexnames = false,citecolor = blue] {hyperref}

\usepackage{soul}
\usepackage{epstopdf}
\usepackage{ulem}

\begin{document}


\title{Electronic properties of \CFS\ investigated by x-ray magnetic linear dichroism}
\author{M. Emmel}
\affiliation{Institut f\"ur Physik, Johannes Gutenberg-Universit\"at Mainz, Staudinger Weg 7, D-55128 Mainz, Germany}
\author{A. Alfonsov}
\affiliation{IFW Dresden, Institute for Solid State Research, D-01069 Dresden, Germany}
\author{D. Legut}
\affiliation{VSB - Technical University of Ostrava, 17.\ Listopadu 15, CZ-70833 Ostrava, Czech Republic}
\author{A. Kehlberger}
\affiliation{Institut f\"ur Physik, Johannes Gutenberg-Universit\"at Mainz, Staudinger Weg 7, D-55128 Mainz, Germany}
\author{E. Vilanova}
\affiliation{Institut f\"ur Physik, Johannes Gutenberg-Universit\"at Mainz, Staudinger Weg 7, D-55128 Mainz, Germany}
\author{I. P. Krug}
\affiliation{Peter Gr\"unberg Institut PGI-6, Forschungszentrum J\"ulich, D-52425 J\"ulich, Germany}
\author{D. M. Gottlob}
\affiliation{Peter Gr\"unberg Institut PGI-6, Forschungszentrum J\"ulich, D-52425 J\"ulich, Germany}
\affiliation{Fakult\"at f\"ur Physik and Center for Nanointegration Duisburg-Essen (CeNIDE), D-47048 Duisburg, Germany}
\author{M. Belesi}
\affiliation{IFW Dresden, Institute for Solid State Research, D-01069 Dresden, Germany}
\author{B. B\"uchner}
\affiliation{IFW Dresden, Institute for Solid State Research, D-01069 Dresden, Germany}
\affiliation{Institut f\"ur Physik, TU Dresden, D-01062 Dresden, Germany}
\author{M. Kl\"aui}
\affiliation{Institut f\"ur Physik, Johannes Gutenberg-Universit\"at Mainz, Staudinger Weg 7, D-55128 Mainz, Germany}
\author{P. M. Oppeneer}
\affiliation{Department of Physics and Astronomy, Uppsala University, P.\,O.\ Box 516, S-751 20 Uppsala, Sweden}
\author{S. Wurmehl}
\affiliation{IFW Dresden, Institute for Solid State Research, D-01069 Dresden, Germany}
\affiliation{Institut f\"ur Physik, TU Dresden, D-01062 Dresden, Germany}
\author{H. J. Elmers}
\affiliation{Institut f\"ur Physik, Johannes Gutenberg-Universit\"at Mainz, Staudinger Weg 7, D-55128 Mainz, Germany}
\author{G. Jakob}
\affiliation{Institut f\"ur Physik, Johannes Gutenberg-Universit\"at Mainz, Staudinger Weg 7, D-55128 Mainz, Germany}

\date{\today}

\newcommand{\low}[1]{\protect\raisebox{-2pt}{\footnotesize{#1}}}
\newcommand{\high}[1]{\protect\raisebox{2pt}{\footnotesize{#1}}}
\newcommand{\textdot}{\textperiodcentered}

\renewcommand{\figurename}{Fig.}
\renewcommand{\tablename}{Table}

\newcommand{\CFS}{Co$_{2}$FeSi}
\newcommand{\CFSA}{Co$_2$FeSi$_{0.6}$Al$_{0.4}$}

\begin{abstract}
We present experimental XMLD spectra measured on epitaxial (001)-oriented thin \CFS\ films, which are rich in features and depend sensitively on the degree of atomic order and interdiffusion from capping layers. Al- and Cr-capped films with different degrees of atomic order were prepared by DC magnetron sputtering by varying the deposition temperatures. The local structural properties of the film samples were additionally investigated by nuclear magnetic resonance (NMR) measurements. The XMLD spectra of the different samples show clear and uniform trends at the $L_{3,2}$ edges. The Al-capped samples show similar behavior as previous measured XMLD spectra of \CFSA. Thus, we assume that during deposition Al atoms are being implanted into the subsurface of \CFS. Such an interdiffusion is not observed for the corresponding Cr-capped films, which makes Cr the material of choice for capping \CFS\ films. We report stronger XMLD intensities at the $L_{3,2}$ Co and Fe egdes for films with a higher saturation magnetization. Additionally, we compare the spectra with \textit{ab initio} predictions and obtain a reasonably good agreement. Furthermore, we were able to detect an XMCD signal at the Si $L$-edge, indicating the presence of a magnetic moment at the Si atoms.

\end{abstract}

\maketitle

\section{Introduction}

Materials with a high spin polarization at the Fermi edge are promising for spintronic devices.\cite{Wolf} Heusler \mbox{compounds}, which have been predicted to be half-metallic ferromagnets, are currently considered as potentially interesting materials for this purpose.\cite{Groot}
Heusler compounds have a distinctive stoichiometry and crystallize in the L2$_1$-structure X$_2$YZ, where X and Y are transition metals and Z is a main group element.
Among the Heusler compounds, \CFS\ has shown the highest magnetization of 6\,$\mu_B$ per formula unit,\cite{Wurmehl} in agreement with the Slater-Pauling rule $m = N_{v} - 24$, where $N_{v}$ is the number of valence electrons.\cite{Fecher}
Moreover, \CFS\ is the Heusler compound with the highest Curie temperature. 
Therefore, \CFS\ might be an ideal material for spintronics applications, provided that the prediction of half-metallic ferromagnetic properties is correct, as was indeed recently suggested by Ref.\,\onlinecite{Bombor}.

However, Co$_2$FeSi is predicted to be a half-metallic compound only if electron correlations are considered in the form of the on-site repulsion energy \textit{U}. Neglecting the electron
correlations, \CFS\ is predicted to be a conventional ferromagnet.\cite{Kallmayer} Hence, a reliable probe of half-metallic ferromagnetism in \CFS\ is still needed. The calculations reported in Ref.\,\onlinecite{Kallmayer} show a significant change in the x-ray magnetic linear dichroism (XMLD) spectra if electron correlation effects are considered. 
Thus, XMLD may serve as a crucial test for band structure calculation schemes as the XMLD, in a single-particle picture, is roughly speaking proportional to the first derivative of the minority density-of-states function (PDOS) in contrast to the x-ray magnetic
circular dichroism being directly proportional to the DOS.\cite{Kunes}
Meinert \textit{et al.}\cite{Meinert} already investigated XMLD for \CFS\ and compared it to a variety of density functional theory-based calculations.
Although the comparison, shown in Ref.\,\onlinecite{Meinert}, rules out some band structure models, qualitative differences between experiment and theoretical predictions still remained.

In this article, we present XMLD spectra measured on a series of \CFS\ film samples with systematically varied atomic order. The increase of atomic order results in an increase of magnetization, which suggests a concomitant transition from metallic to half-metallic ferromagnetic behavior. Thus, according to theoretical calculations, a considerable altering of the spectra is expected. To address the relation between the measured XMLD spectra and the underlying electronic structure, we employed various exchange-correlation functionals, used a few different values of the Hubbard $U$ ($U_{\rm eff}$) parameter and performed a number of calculations using the so-called non-selfconsistent fixed spin moment technique. Although the overall spectrum is well given, the fine structure appearing predominantly at the $L_3$ edge is not sufficiently captured.

\section{Experimental}

Epitaxial (001)-oriented \CFS\ films were grown by DC magnetron sputtering onto MgO(001) substrates in a UHV system with a base pressure of 1\textperiodcentered 10$^{-9}$$\,$mbar. The distance between the film and the stoichiometric target of \CFS\ was 10\,cm, while the Ar working pressure was 7.77\textperiodcentered 10$^{-2}$$\,$mbar. Different degrees of local atomic order in the films were achieved by setting different substrate temperatures at each film deposition in the range of 470$^\circ$C to 730$^\circ$C. The substrate temperature was determined with a pyrometer set to an emissivity of 0.31. To prevent oxidation the films were capped immediately after deposition with a 3\,nm thick capping layer grown at room temperature. We used both Al and Cr as a capping layer.

The saturation magnetization of each film was determined with a superconducting quantum interference device (SQUID) magnetometer (Quantum Design, MPMS-XL-5) at 20\,K and 300\,K with magnetic fields up to 3\,T.
Thickness determination and structural characterization were performed by x-ray reflection/diffraction with a Phillips X'pert diffractometer.

The x-ray absorption spectra (XAS) were recorded at the synchrotron light source BESSY II (beamline UE56/1-SGM).
The samples were mounted perpendicular to the beam and were magnetically saturated by an external field parallel to the film surface.
The degree of linear polarization of the synchrotron radiation was close to 100\,\%. All measurements were performed at room temperature with the total-electron-yield (TEY) technique where the specimen current is measured. The photon intensity was measured by a gold mesh.
The polarization of the photon beam was switched between linear vertical and linear horizontal while the magnetic field was kept constant. The linear dichroism is calculated as the difference between spectra measured with horizontal and vertical polarization.

The NMR experiments were performed in an automated, coherent, phase
sensitive and frequency tuned spin-echo spectrometer provided by NMR
Service Erfurt, Germany. The NMR spectra were recorded at a temperature of 5\,K and
in the frequency range of 120-200\,MHz with steps of 0.5\,MHz. No
external magnetic field was applied during the measurements. All NMR spectra were
corrected for the magnetic enhancement factor as well as the $\omega^2$
dependence, resulting in relative spin-echo intensities, which are proportional to
the number of nuclei with a given NMR resonance frequency. The procedure of the
enhancement correction is described in Refs.\,\onlinecite{Wurmehl2008review} and \onlinecite{pan97}.

\subsection{\textit{Ab initio} calculations}

The electronic structure of the \CFS\ was calculated with the WIEN2k code\cite{Wien2k} in the density-functional theory (DFT) framework, employing the local spin density approximation\cite{LSDA} and the general gradient appoximation (GGA), the latter as parameterized by Perdew-Burke-Ernzerhof\cite{PBE} for the exchange-correlation term. Electronic correlations were included by using the so-called Hubbard \textit{U} approach (LSDA+\textit{U}, GGA+\textit{U}) in the self-interaction corrected formulation.\cite{Anisimov93} Calculations on the basis of the conventional usual local spin density approximation or the GGA alone, i.e. without additional Hubbard \textit{U}, do not give a correct magnetic moment for Co$_2$FeSi as was already pointed out before by several authors (see Refs.\,\onlinecite{Kallmayer} and \onlinecite{Meinert} and references therein). All calculations presented here were done for the lattice parameter $a = 5.64$\,\AA.  

The core electrons were treated fully relativistically (Dirac equations) and for the semicore and valence states the spin-orbit interaction (SOI) was included through a variational scheme.\cite{Wien2k} The convergence of the electronic structure calculations was ensured by the following parameters: the convergence criterion of the total energy was better than $10^{-7}$ Ry/atom; the energy cutoff given as the product of the muffin-tin radius and the maximum reciprocal space vector $R_\mathrm{MT}K_\mathrm{max}$ was 8.5, the largest reciprocal vector in the charge Fourier expansion, $G_\mathrm{max}$, was set to 12\,Ry$^{1/2}$, the maximum angular momentum value of partial waves inside the muffin-tin spheres $l_\mathrm{max}$ was 10, and a grid of 20$\times$20$\times$20 $k$-points was applied to sample the 1$^{\rm st}$ Brillouin zone. The same $k$-point mesh was used to determine dipolar squared momentum matrix elements, which occur in the Kubo formula for the permittivity tensor. To account for lifetime effects a broadening with a Lorentzian with a width of 0.3\,eV was applied for the x-ray spectra from core (localized) to unoccupied states (delocalized states). The core level exchange splitting was also included.\cite{Kunes03} The Kramers-Kronig transformations were performed and the complex values of permittivity tensor elements $\epsilon_{ij}$ were obtained. Subsequently, permittivity elements served as input for our in-house optical code solving Fresnel equations using 4-vector Yeh's formalism as extended for magnetic multilayers.\cite{Vis91} We hence obtain the reflectivity and transmittivity matrices ($R_{pp},R_{ss},R_{ps},R_{sp},T_{pp},T_{ss},T_{ps},T_{sp}$) for each magnetization direction \textbf{M}. The absorbance $A_b$ is defined as 
\begin{equation}
A_b = -\mathrm{ln}(T) = -\mathrm{ln}(1-A-S-R),
\label{eq:absorbance}
\end{equation}
where $A$ is absorptance, $S$ denotes scatter and $R$ reflection of the material. The thickness dependent reflectivity and transmittivity matrices then lead 
to the attenuation coefficient $\mu$, which for both polarizations $E_1$ $\|$ $\textbf{M}$ and $E_2 \perp$ $\textbf{M}$ is calculated according to the Beer-Lambert law
\begin{equation}
I(d) = I_{0}e^{-\mu_{p}(\omega)d},
\label{eq:beers law}
\end{equation}
where the incident intensity $I_0 = 1$ and the attenuated intensity for the first (second) polarization $P = E_1~(E_2)$ is $I_{pp(ss)}(d) = [T_{pp(ss)}-R_{pp(ss)}]/d$, where $d$ is the sample thickness. Note, that the absorption coefficient $\mu$ differs from absorptance $A$ defined as $A = \frac{I_0 - I}{I_0}$, i.e. incoming and absorbed intensity. Finally, the linear dichroism is computed here as XMLD = $(\mu(I_{pp})-\mu(I_{ss}))/norm$. Note, that this holds only for normal incidence. The normalization factor used was $norm = max(\frac{\mu_{E_1}+\mu_{E_2}}{2})$.

\section{Structural properties}

The long range order of the atomic structure of the films was determined with 4-circle x-ray diffraction. The obtained lattice constant of the films is 5.64 $\pm$ 0.01\,\AA, in agreement with previous measurements of \CFS\ thin films as well as bulk material.\cite{Schneider} The rocking curves reveal widths $\Delta \omega$ between 0.29$^\circ$ and 0.73$^\circ$. To determine the L2$_1$ structure within the sample series, it is not sufficient to compare the maximum intensities of the (202) - and (111) - film reflections. Rather the ratio of the integrated peak areas of $\omega$ - scans $A_{202}/A_{111}$ provides a better estimation of the real intensity ratios. We measured the rocking curves of the (202) - and (111) - reflections, which were fitted with Gaussian functions to determine the enclosed areas of the reflections. We compare our values with results calculated for powder diffraction corrected with the geometry factor according to Bragg Brenato and multiplicity. The ratio of the individual reflections is given by the program \textit{PowderCell} as $I_{202} /I_{111} = 12.08$. Higher ratios of $A_{202}/A_{111}$ are due to Fe atoms located on Si sites and vice versa.
Structural data and magnetization values measured by a SQUID magnetometer, as well as NMR results, are summarized in Table \ref{tab:samples}.
\begin{table*}[ht]
  \centering
    \caption{Deposition temperature, thickness of capping layer, saturation magnetization at 20\,K and width of the rocking curve of the 400-film reflex, area ratio $A_{220}$/$A_{111}$, and NMR parameters for the epitaxial Co$_2$FeSi/MgO films.}
  \label{tab:samples}
  \begin{tabular}{cccccccccc}
   \hline
	\hline
    Film & T$_{Sub}$  & d$_{capping}$ &  m$_{sat}$ & $\Delta \omega$ & $I_{202}/I_{111}$  & NMR freq. & NMR linewidth & NMR spacing & Fe-Si off-\\
         &(\,$^{\circ}$C)& (nm) & ($\mu_B$/f.u.) &  ($\,^{\circ}$) & & (MHz) & (MHz) & (MHz) &  stoichiometry (\%)\\
      \hline
            
    CFS(Cr1) & 470 & 3.0 Cr& 4.89 & 0.87 & 21.7 & 144.85$\pm0.03$ & 8.59$\pm0.07$ & 33$\pm1$ & 1.59$\pm0.19$  \\        
    CFS(Cr2) & 620 & 3.0 Cr& 4.33 & 0.55 & 17.0 & 145.54$\pm0.03$ & 8.03$\pm0.07$ & 33$\pm2$ & 1.38$\pm0.39$  \\        
    CFS(Cr3) & 640 & 3.0 Cr& 5.37 & 0.46 & 15.1 & 144.41$\pm0.03$ & 8.03$\pm0.07$ & 31$\pm1$ & 1.12$\pm0.20$  \\        
    CFS(Cr4) & 730 & 3.0 Cr& 5.80 & 0.37 & 11.1 & 144.15$\pm0.03$ & 7.90$\pm0.07$ & 27$\pm3$ & 0.32$\pm0.20$  \\        
    \hline
    CFS(Al1) & 540 &  3.0 Al& 5.02 & 0.54 & - & 144.89$\pm0.03$ & 8.39$\pm0.07$ & 33$\pm1$ & 1.72$\pm0.17$  \\            
    CFS(Al2) & 640 &  3.0 Al& 5.31 & 0.69 & - &144.42$\pm0.03$ & 7.68$\pm0.07$ & 30$\pm1$ & 1.38$\pm0.48$  \\             
    CFS(Al3) & 715 &  3.0 Al& 4.98 & 0.35 & - &144.68$\pm0.05$ & 8.75$\pm0.14$ & 31$\pm1$ & 1.48$\pm0.22$  \\             
    CFS(Al4) & 715 &  3.6 Al& 6.10 & 0.34 & - &146.60$\pm2.00$ & 6.57$\pm0.50$ & 28$\pm4$ & 0.47$\pm0.65$  \\            
     \hline
    \hline
\end{tabular}
\end{table*}
We see an increase of the saturation magnetization with higher substrate temperatures, except for CFS(Cr2), which exhibits the lowest value. In the Al-capped series we find the same for CFS(Al3).
In general a systematic dependency for all values on the substrate temperature is observed.

\begin{figure*}[htbp]
\vspace*{0.0cm}
\includegraphics[width=2.00\columnwidth]{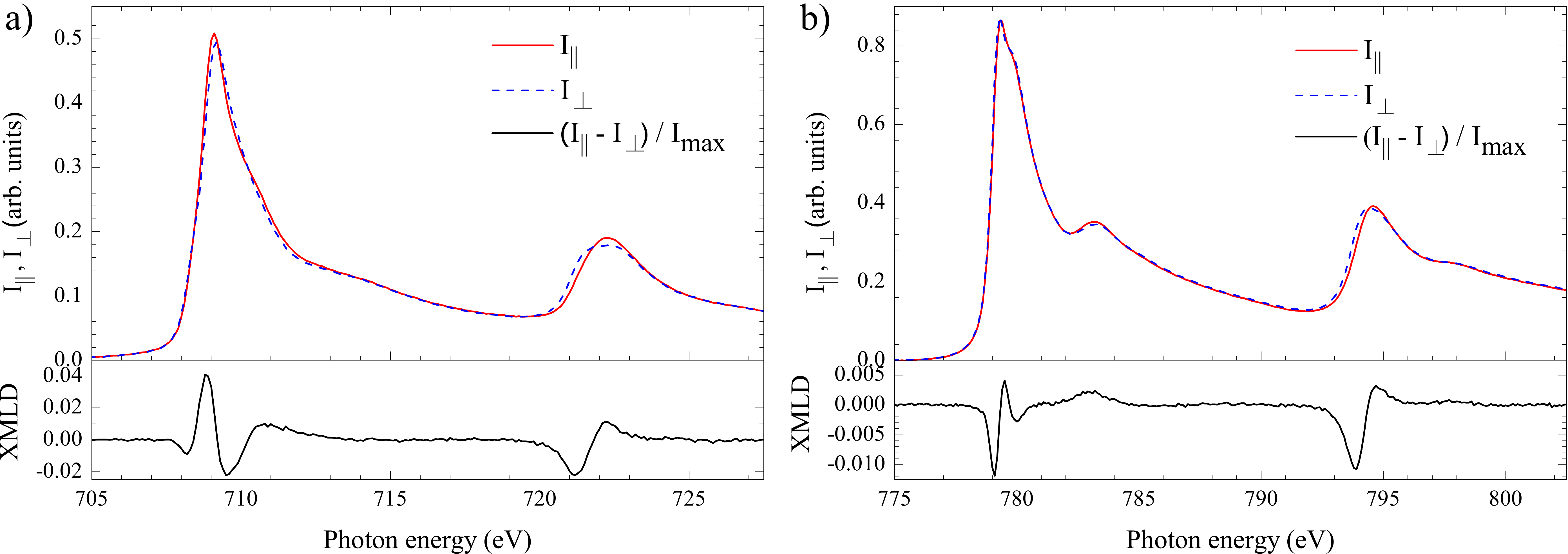} 
\hspace*{0cm}
\caption{\label{fig:XASandXMLD} (Color online) X-ray absorption spectra measured on the \CFS (Cr4) sample at the $L_{3,2}$ edge of Fe (a) and Co (b) with the polarization vector parallel ($I_{\|}$) and perpendicular ($I_{\bot}$) to the film magnetization along the [100] direction. The resulting XMLD spectra \mbox{$(I_{\|} - I_{\bot})$/$I_{max}$} are shown in the lower panels, where $I_{max}$ is the maximum value of the corresponding mean $L_{3}$ absorption edge.}
\end{figure*}

\section{Results}
\label{sec:Results}

\subsection{Experimental results}

\subsubsection{XMLD spectra}
The absorption intensity is measured for the electric field vector of the linearly polarized beam parallel $I_{\|}$ and perpendicular $I_{\bot}$ to the magnetization vector. The differences $I_{\|} - I_{\bot}$ result in the corresponding XMLD spectra. The XMLD spectra were measured for each sample at the Co and Fe $L_{3,2}$ edges with the magnetization vector along the in-plane directions [100] and [110]. Each XMLD spectrum was normalized by the maximum value of the respective $L_3$ edge $I_{max}$ of the averaged XAS spectrum.
Fig.\,\ref{fig:XASandXMLD} shows exemplarily two XAS spectra measured in [100] orientation at the Fe (a) and Co (b) edge with the corresponding XMLD spectra.
At the Co [100] XAS spectra the maximum at the $L_3$ edge is followed by a shoulder with a distance of 4\,eV. This is typical for the Heusler structure and was already observed for different Heusler compounds (Refs.\,\onlinecite{Telling,Klaer,Elmers}). An analogous shoulder is visible for the Co edge at the energy 798.1\,eV above the $L_2$ peak, which is smaller due to lifetime broadening effects.\cite{Elmers}

\begin{figure*}
\vspace*{0.0cm}
\includegraphics[width=1.95\columnwidth]{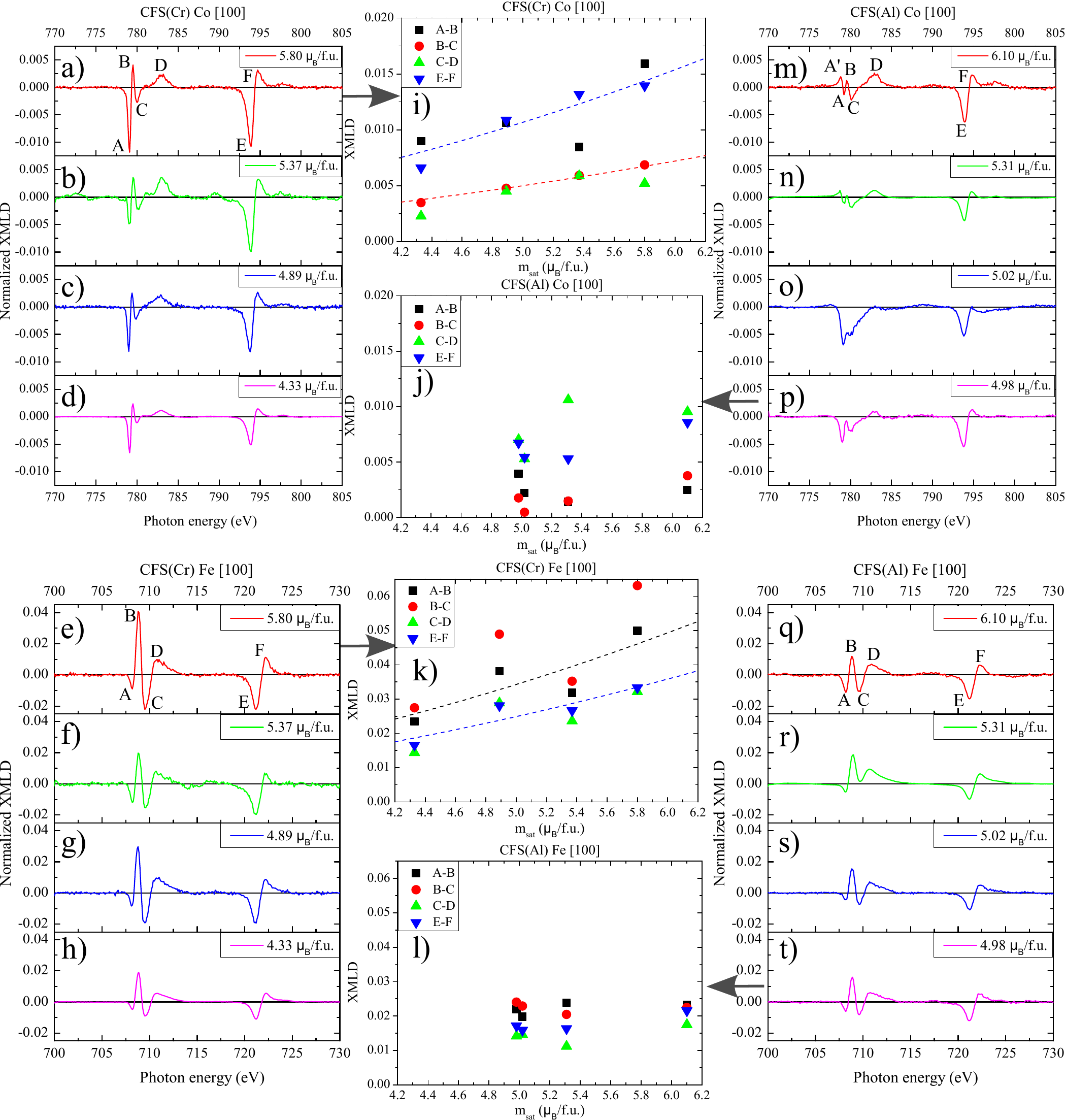} 
\hspace*{0cm}
\caption{\label{fig:overview100} (Color online) X-ray magnetic linear dichroism spectra measured on the \CFS\ films with Cr capping layer (a - h) and Al capping layer (m - t) taken at [100] orientation. First row: Results with photon energies at the Co edge. Second row: Results with photon energies at the Fe edge. The saturation magnetization of each film is displayed in the upper right corner. Middle row: Four graphs (i - l), each related to a series of XMLD spectra (indicated by the arrows), showing the strength of the XMLD depending on the magnetization of the film. In the legend A-B, C-D, E-F and E-F relate to the difference of the XMLD signal between the respective points. Colored dashed lines represent a quadratic fit of the XMLD values of the same color.}
\end{figure*}

\begin{figure*}
\vspace*{0.0cm}
\includegraphics[width=1.95\columnwidth]{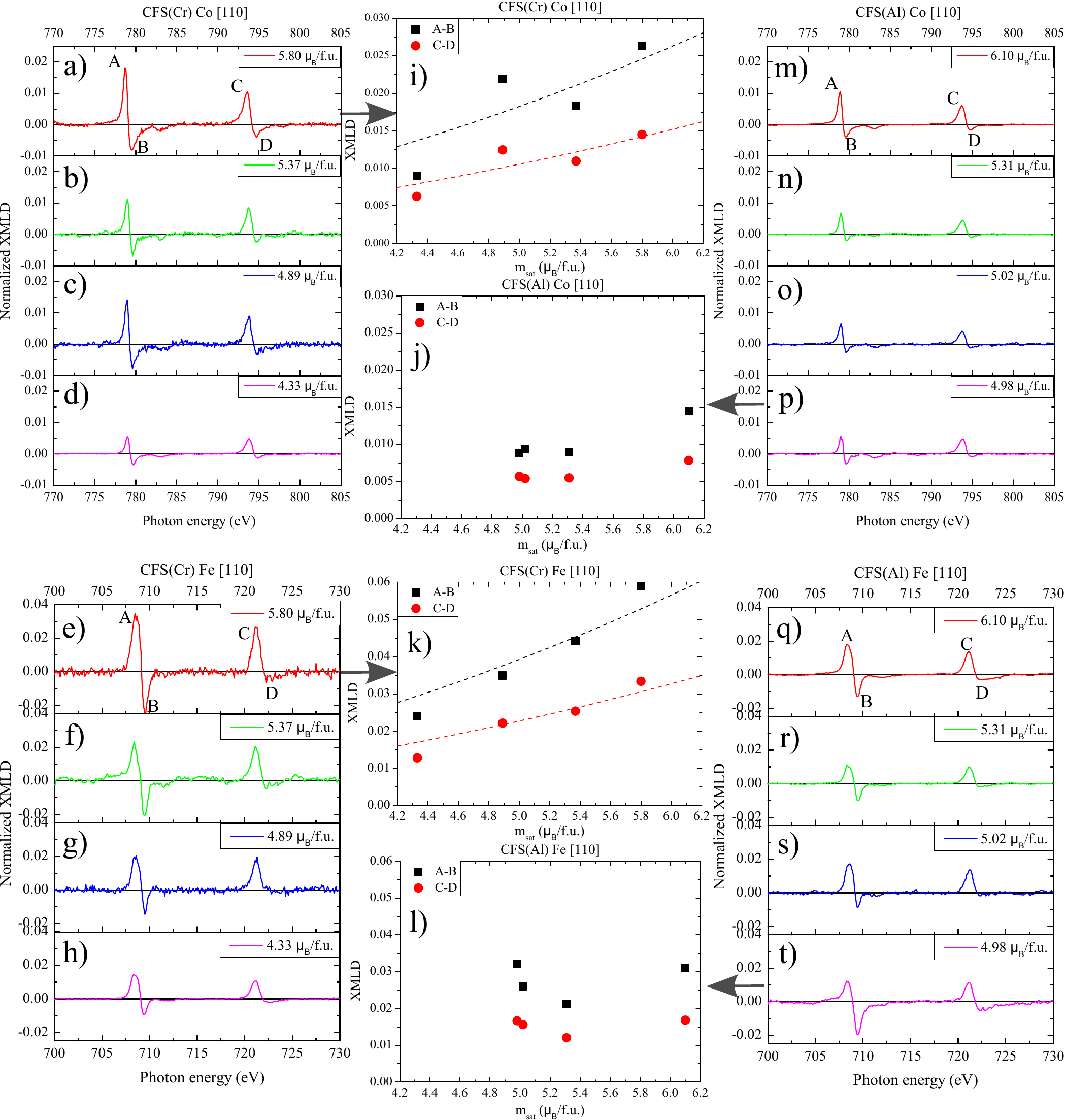} 
\hspace*{0cm}
\caption{\label{fig:overview110} (Color online) X-ray magnetic linear dichroism spectra measured on the \CFS\ films with Cr capping layer (a - h) and Al capping layer (m-t) taken at [110] orientation. First row: Results with photon energies at the Co edge. Second row: Results with photon energies at the Fe edge. The saturation magnetization of each film is displayed in the upper right corner. Middle row: Four graphs (i - l), each related to a series of XMLD spectra (indicated by the arrows), showing the strength of the XMLD depending on the magnetization of the film. In the legend A-B and C-D relate to the difference of the XMLD signal between the respective points. Colored dashed lines represent a quadratic fit of the XMLD values of the same color.}
\end{figure*}

The complete set of XMLD spectra for each film is presented in Fig.\,\ref{fig:overview100} (for the case of $\vec{M}$ $\|$ [100]) and Fig.\,\ref{fig:overview110} (for the case of $\vec{M}$ $\|$ [110]).
For direct comparison, the XMLD spectra for Cr- and Al-capped films are displayed on the left- and right-hand side, respectively. The spectra measured at the Co $L$-edge are presented in the top panels and those measured at Fe $L$-edge in the bottom panels.

The structure of the XMLD spectra clearly exhibits more features at the $L_{3,2}$ edge than the XMCD spectra of \CFS, which was already predicted by Ref.\,\onlinecite{Kallmayer} and reported in Ref.\,\onlinecite{Meinert}. The spectra measured for the \CFS (Cr4) film with the deposition temperature of 730$^\circ$C (Fig.\,\ref{fig:XASandXMLD} and Fig.\,\ref{fig:overview100}\,a,e) agree in all details of the spectra with the spectra reported in Ref.\,\onlinecite{Meinert}.
The largest XMLD signal is observed for the \CFS (Cr) film with the highest magnetization (5.8\,$\mu_B$/f.u.). Maximum XMLD values of 4.1\,\% at the Fe $L_3$ edge for $\vec{M}$ $\|$ [100] (Fig.\,\ref{fig:overview100}\,e) and 3.4\,\% at the Co $L_3$ edge for $\vec{M}$ $\|$ [110] (Fig.\,\ref{fig:overview110}\,e) are obtained.


\subsubsection{Influence of the capping layer}
\label{subsec:capping}
Due to the surface sensitivity of the TEY method different interface properties for different capping materials can be detected.
The information depth of TEY is given by the mean free path of the emitted secondary electrons and typically amounts to 2.5\,nm.\cite{Kallmayer} We compare two series of samples prepared with identical conditions except for the different capping materials Al and Cr.
Both capping materials do not contribute to the absorption spectra in the investigated energy range. Thus, we anticipated no change in the measured XMLD spectra of the films with different capping layers.
Contrarily, the two sets of spectra show pronounced differences, as discussed in the following.

The XMLD spectra of the \CFS(Cr) films (see left column of Fig.\,\ref{fig:overview100} and Fig.\,\ref{fig:overview110}) show similar features within the film series at a given field and beam geometry.
In contrast, the XMLD spectra of Al-capped films (see right column of Fig.\,\ref{fig:overview100} and Fig.\,\ref{fig:overview110}) show pronounced variations for different samples.

The most striking differences between Cr- and Al-capped films are observed for $\vec{M}$ $\|$ [100] at the Co edge (Fig.\,\ref{fig:overview100}\,a and Fig.\,\ref{fig:overview100}\,m). For the \CFS(Cr) series at the $L_3$ edge, we note a sharp minimum (A) followed by a sharp maximum (B).
For the \CFS(Al) samples, instead, the XMLD spectra starts with a maximum (A') followed by the minimum A (Fig.\,\ref{fig:overview100}\,m,n). Samples with lower magnetization (Fig.\,\ref{fig:overview100}\,o,p) show a minor (A) but do not show a positive maximum (B).

\begin{figure*}
\vspace*{0.0cm}
\includegraphics[width=1.95\columnwidth]{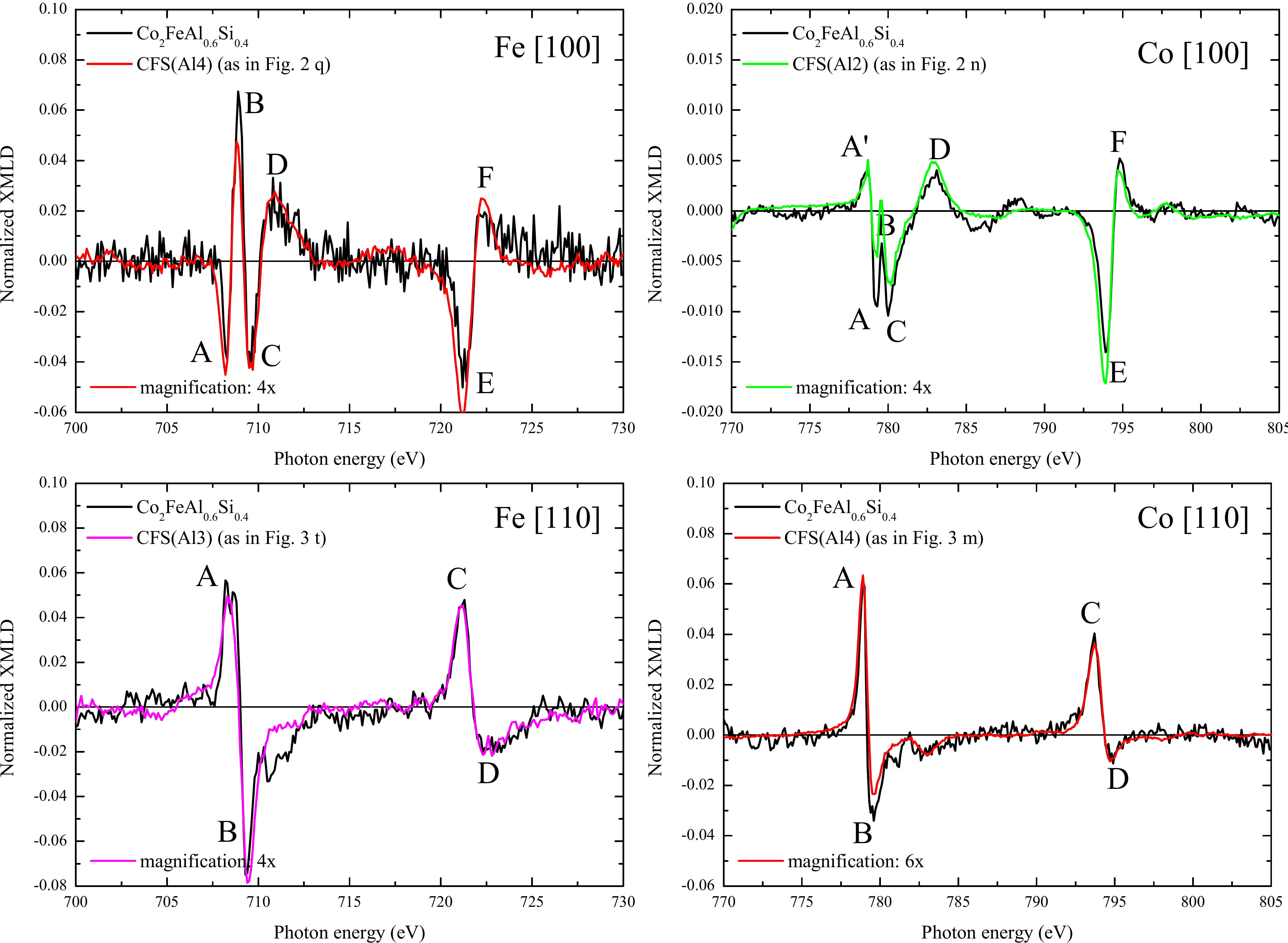} 
\hspace*{0cm}
\caption{\label{fig:CFSAaverage} (Color online) X-ray magnetization linear dichroism measured on a \CFSA\ film, capped with 3\,nm Al layer, at the $L_{3,2}$ edge of Fe and Co with the film magnetization along the [100] and [110] direction. Each XMLD spectrum is compared with a spectrum measured on one of the \CFS(Al) films.}
\end{figure*}

We assume that the Al atoms interdiffuse into the film. To support our assumption we recorded XMLD spectra (Fig.\,\ref{fig:CFSAaverage}) of a \CFSA\ film, which was deposited at 700$^\circ$C and capped with a 3\,nm Al layer. \CFSA\ was chosen to mimic a \CFS\ film with an interdiffusion of Al atoms in the subsurface for the surface-sensitive TEY method. The spectra reveal the same features as for the Al-capped \CFS\ films for both Fe and Co and directions [100] and [110].
Especially at Co and for direction [100] we see that a maximum A$^{\prime}$ is followed by a minimum A, like discussed before. Given the similarity of the XMLD spectra of the \CFSA\ film and the \CFS(Al) series, we assume that during the deposition of the capping layer, Al atoms interdiffuse into the \CFS\ film. Even though we grew each capping layer at room temperature, our results suggest intermixing, also being partially due to resputtering of the \CFS\ atoms with the result of Al atoms being implanted into the first atomic layers. As a consequence, the subsurface of the prepared \CFS(Al) films is similar to \CFSA. Since the TEY method is surface sensitive, the measured XMLD on the \CFS(Al) films resemble the XMLD spectra of the \CFSA\ film. The interdiffusion of the Al atoms manifests in each XMLD spectra differently with respect to the orientation (i.e.\ the excitation geometry which the sample was measured in).
Additionally, the probed element (Co, Fe) plays a role, as the influence of the Al atoms in the upper layer can be clearly observed at the Co edge, whereas the capping has hardly any influence on the XLMD spectra of Fe edge. This is easily understood by taking into account the following argument: The Al capping atoms intermix with Si atoms. This interdiffusion changes the first nearest neighboring shell of 
Co atoms and only the second nearest neighbor shell of Fe.
The non systematic variation can be explained by unknown temperatures of the sample during the deposition of the capping layer. This took place a temperatures between room temperature and 50$^\circ$C, where we did not expect any interdiffusion and therefore did not monitor exact temperatures and waiting times. Thus different degrees of intermixing of Al atoms with the film occur.
In general it is possible to show interdiffusion via neutron reflectivity measurements of multilayer sample (see Ref. \onlinecite{Zabel}). However, to show interdiffusion on a single layer is much more demanding due to the small sample volume and the unknown partial oxidation of the capping layer.

As mentioned above, we encounter no large variations for the Cr-capped films. Hence, we conclude that Cr does not interdiffuse with the \CFS\ film.
A non continuous capping layer would result in a partially oxidized film, which clearly would change the XMLD spectra. We can exclude this as we already showed XMCD spectra of samples without oxidation in Ref. \onlinecite{Kallmayer2} that were capped with the same material and procedure.

\subsubsection{XMLD versus Magnetization}
\label{subsec:XMLDvsMagnet}

The XMLD is expected to increase with the square of the magnetization of the sample. To investigate this dependence systematically, the intensity differences of the marked maxima and minima (A, B, C, D, E, and F) were taken as a scale for the XMLD strength. The respective XMLD values were plotted versus the corresponding saturation magnetization of the film (Fig.\,\ref{fig:overview100}\,i,j,k,l and Fig.\,\ref{fig:overview110}\,i,j,k,l).
The Co[100]-plot (Fig.\,\ref{fig:overview100}\,i) of the Cr-capped series exhibits an increase of B-C and E-F. The value of A-B of the film with m$_{sat}$ = 5.37\,$\mu_B$/f.u. does not match the general trend of an increase, even though the other samples depict an increase. C-D shows stronger XMLD values with rising saturation magnetization, except the last datapoint of C-D, which is slightly lower than the film with with m$_{sat}$ = 5.37\,$\mu_B$/f.u..
However, a trend of increasing XMLD signal is visible, whereas at the Co[100]-plot for the Al-capped series (Fig.\,\ref{fig:overview100}\,j), no such dependence can be observed.
For the Fe[100]-plot with the Cr-capped series (Fig.\,\ref{fig:overview100}\,k), the film with the lowest and the highest saturation magnetization have the smallest, respectively the highest XMLD values as expected. However, the film with 5.37\,$\mu_B$/f.u. shows smaller values than the film with 4.89\,$\mu_B$/f.u.. The same can be observed for Co [110] (Fig.\,\ref{fig:overview110}\,i).
The Al-capped series of the Fe[100]-plot (Fig.\,\ref{fig:overview100}\,l) and Co[110]-plot (Fig.\,\ref{fig:overview110}\,j) do show very similar values for the XMLD and thus again show no dependence on the saturation magnetization.
For the Cr-capped series at the Fe[110]-plot (Fig.\,\ref{fig:overview110}\,k), the value for A-B and C-D increases with the saturation magnetization. For the Al-capped series (Fig.\,\ref{fig:overview110}\,l), it is found that the film with the lowest saturation magnetization of 4.98\,$\mu_B$/f.u. has the strongest XMLD features. Even the film with 6.10\,$\mu_B$/f.u. does not exceed these values.
Generally, comparing the XMLD spectra of the Cr- and Al-capped films, higher values for the Cr-capped film series were observed.

\subsubsection{NMR results}

Nuclear magnetic resonance (NMR) probes all the $^{59}$Co nuclei in the film samples due to a rather large penetration depth of the radio frequency wave (typically on the order of $\mu$m). The surface effects seen in XMLD, namely the formation of a few monolayers of Co$_2$FeSi$_{0.6}$Al$_{0.4}$ at the \CFS/Al interface, are not observable by NMR due to a lack of sensitivity, as the signal was optimized to reveal the NMR signals stemming from the bulk of the films. Note that the NMR spectra of \CFSA\ (at about 160~MHz) may be hidden in the broad signal. NMR allows to investigate the local environments of $^{59}$Co nuclei in the bulk of the film samples, and enables to determine the degree of order as well as possible existence of off-stoichiometry.\cite{Wurmehl2008review,WKS2009} 

\begin{figure}
\includegraphics[width=0.95\columnwidth]{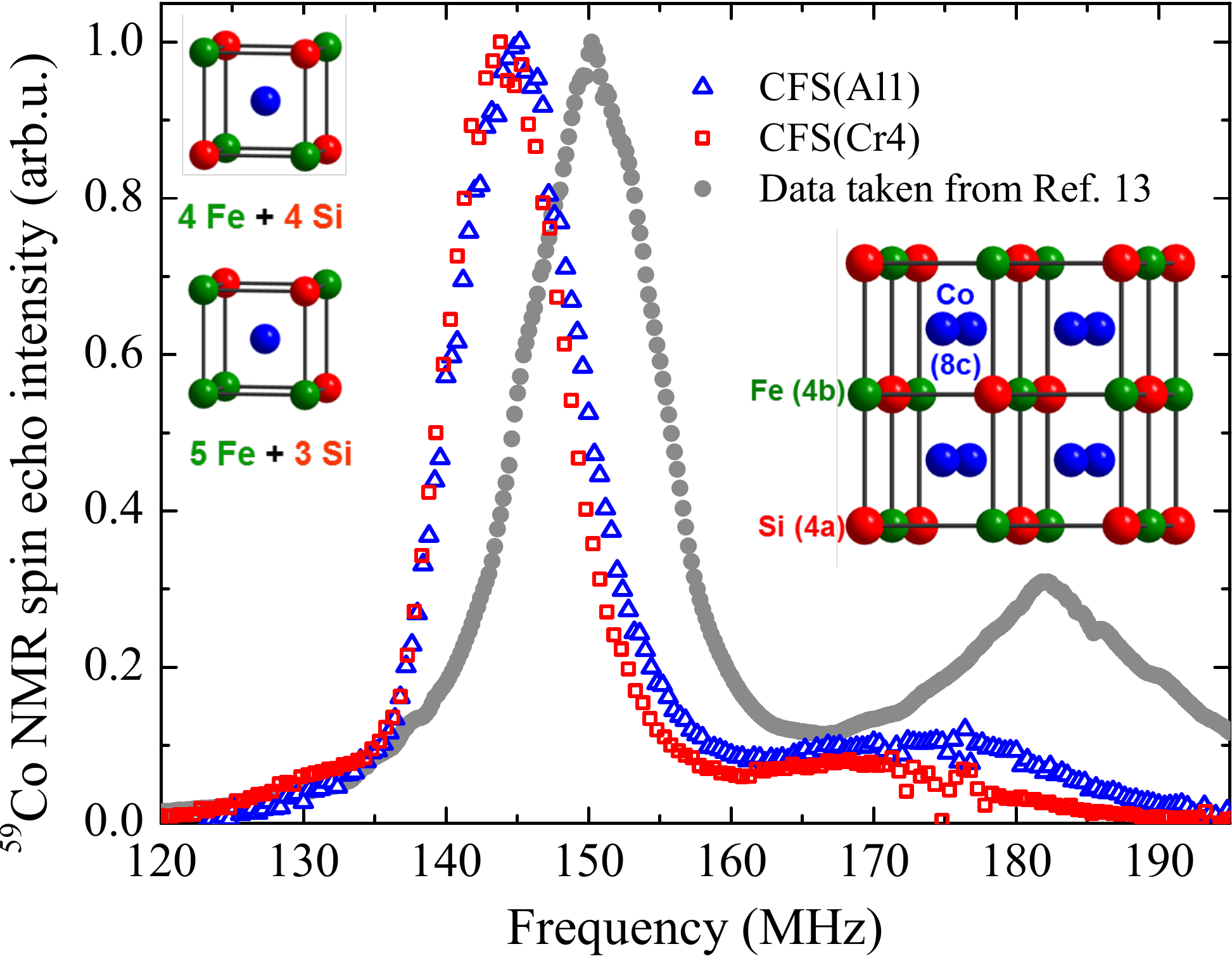}
\hspace*{0cm} \caption{\label{fig:NMR1} (Color online) $^{59}$Co NMR spectra of CFS(Al1) and
CFS(Cr4) samples plotted together with previous data on similar Co$_2$FeSi sample,
data taken from Ref.\,\onlinecite{WKS2009}. The upper inset shows the $L2_1$ type Heusler structure with the corresponding Wyckhoff positions of Co (blue), Fe (green)  and Si (red). Please note that the first coordination shell of Co in the $L2_1$ type structure with the 2:1:1 stoichiometry consists of 4Fe+4Si atoms; the resonance line at 145~MHz corresponds to $^{59}$Co nuclei with this specific local environment. The satellite at higher frequencies originates in an Fe-rich stoichiometry with the first coordination shell of Co consisting of 5Fe+3Si atoms (compare Ref.\,\onlinecite{WKS2009}).}
\end{figure}

The $^{59}$Co NMR spectra measured on all samples (representatives are shown in Fig.\,\ref{fig:NMR1}) have a main line at around 145~MHz with one satellite on the high-frequency side. A similar $^{59}$Co NMR spectrum with high-frequency satellites was reported for off-stoichiometric Co$_2$FeSi films (Ref.\,\onlinecite{WKS2009}). In the present case, the average linewidths (7.99~MHz) are smaller than those reported in the previous data (11.8~MHz, Ref.\,\onlinecite{WKS2009}), which already indicates a higher degree of order and smaller degree of off-stoichiometry in the present films. This interpretation is consistent with the overall shift of the main line (145~MHz) towards lower frequencies compared to previous NMR data on \CFS\ films (150\,MHz\cite{WKS2009}), the main line of the Co$_2$FeSi films now shifting towards the single NMR line observed for highly-ordered bulk Co$_2$FeSi (139~MHz \cite{Blum2009}). The intensity of the high-frequency satellite lines is lower compared to the satellite lines presented in Ref.\,\onlinecite{WKS2009}, indicating that the films investigated here are of much higher structural quality.
The improved short range order of the L2$_1$ structure visible in the NMR parameters correlates with the improved long range order seen in the X-ray investigations, i.e. an increased intensity of the (111) ordering peak.

\begin{figure}
\includegraphics[width=0.95\columnwidth]{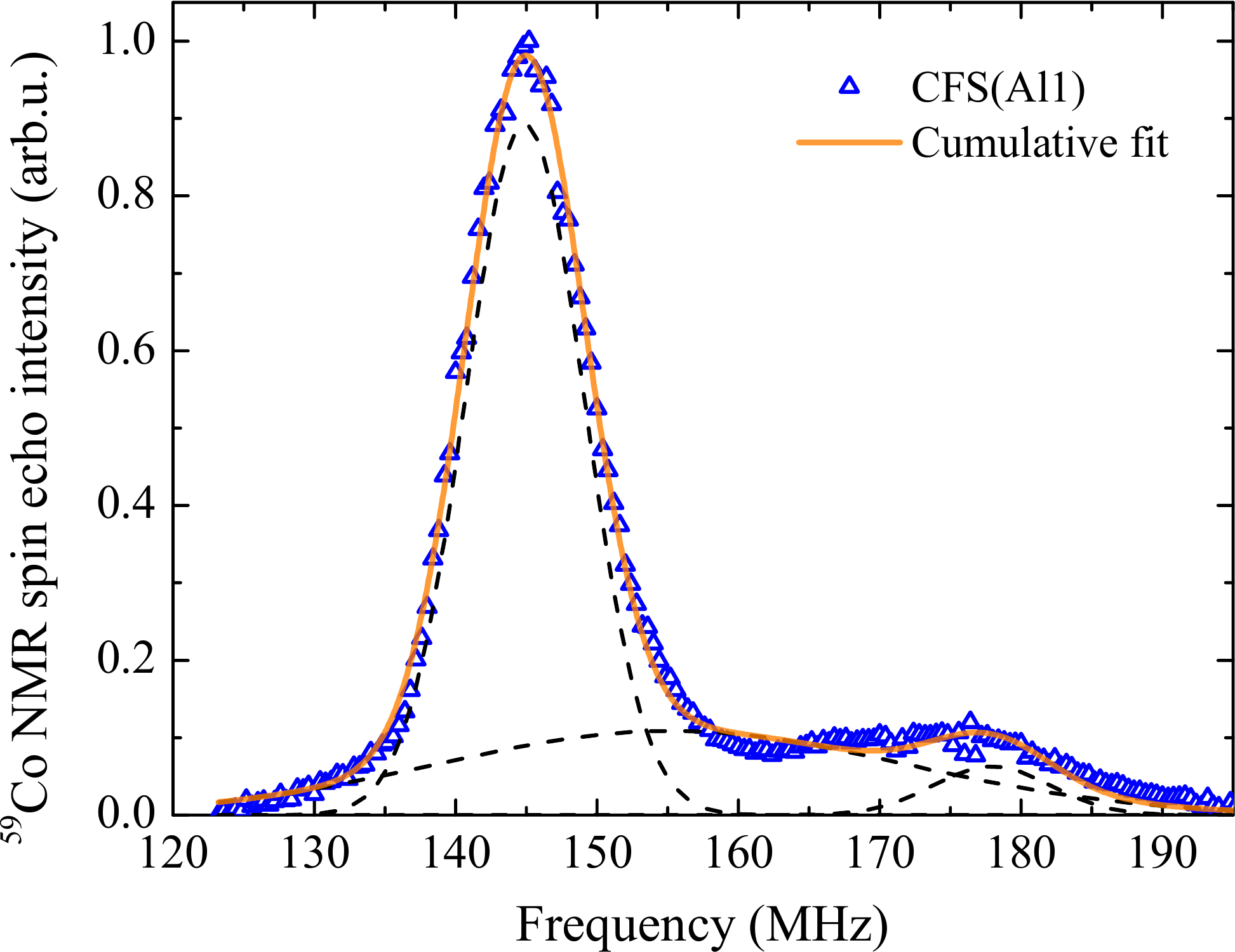}
\hspace*{0cm} \caption{\label{fig:NMR2} (Color online) $^{59}$Co NMR spectrum of \CFS(Al1) sample
and its fit using three Gaussian lines. Thick solid line represents the total fit
of the spectrum.}
\end{figure}

\begin{figure*}
\includegraphics[width=2\columnwidth]{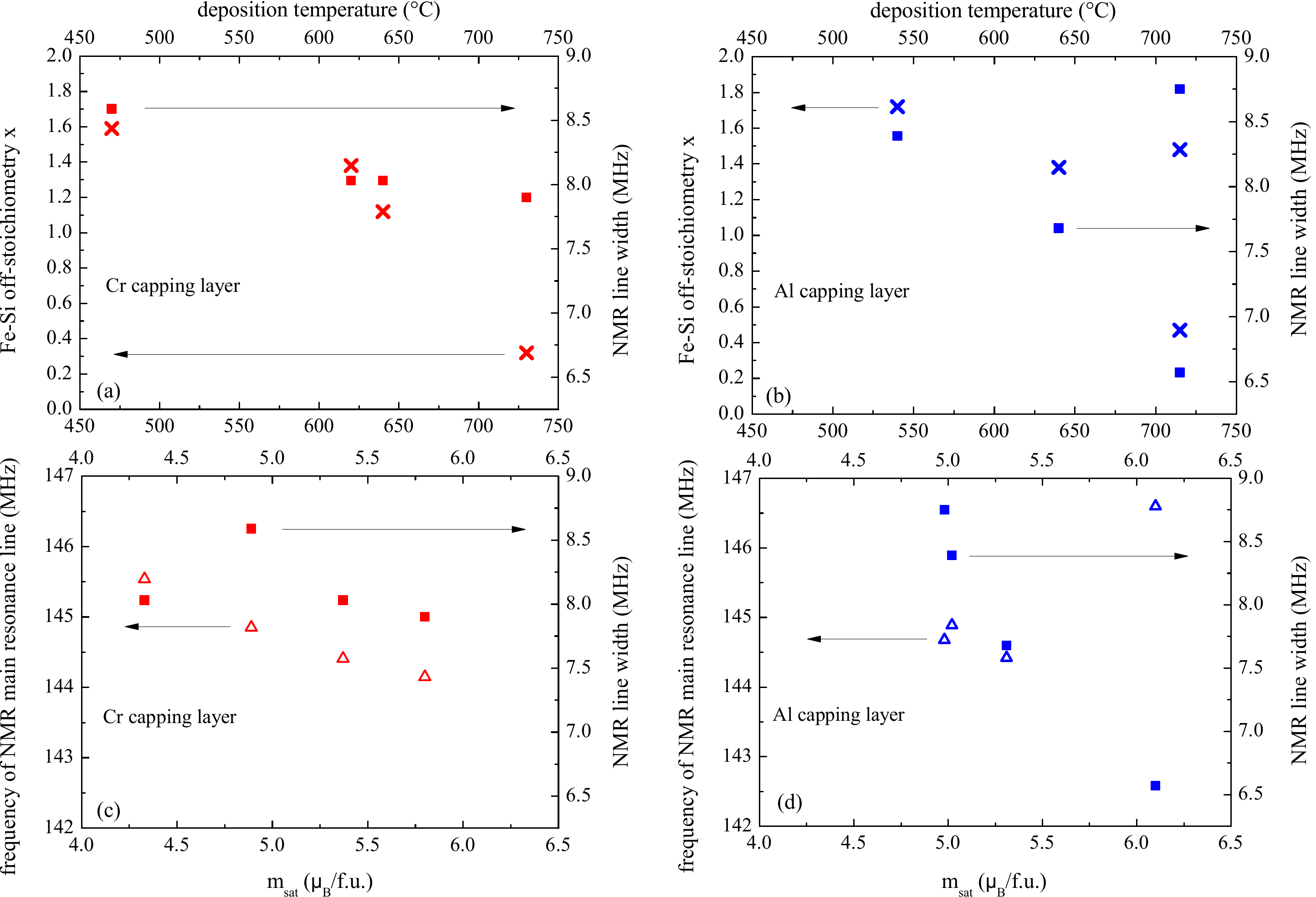}
\hspace*{0cm} \caption{\label{fig:NMR} (Color online) Relation between the results of the NMR analysis (amount of Fe-Si off-stoichiometry $x$ (crosses)), width (squares) and frequency (triangles) of NMR main line) and the film characteristics (deposition temperature (a,b) and moment in saturation (c,d)). The data for the Cr capped films are shown in the left panels (a,c) while the data for the Al capped films are presented in the right panel (b,d). }
\end{figure*}

To quantitatively analyze the spectra we fitted them using a sum of Gaussian lines representing different numbers of Fe and Si next nearest neighbors in the first coordination shell of Co (compare Ref.\,\onlinecite{WKS2009}). The representative fit of the \CFS(Al1) sample is shown in Fig.\,\ref{fig:NMR2}. However, the fit matches much better to the data if a third broad line is included. Interestingly, the resonance frequency of this broad line coincides with the one observed for the main line in Co$_2$FeSi$_{0.5}$Al$_{0.5}$.\cite{Wurmehl2012} However, this line is observed for both types of capping layers, Al and Cr. This excludes the formation of Co$_2$FeSi$_{0.6}$Al$_{0.4}$ as origin of the broad NMR line. We suggest the origin of the broad line to be strain within the film, induced by the substrate as explained in the following: The broad line has the same frequency and the same linewidth for all samples, the only difference is its relative contribution to the total spectrum, which slightly varies from sample to sample. The contribution of the broad line is stronger for thinner films, e.g.\ in the case of the 45\,nm Cr-capped \CFS\ film the ratio between intensities of the broad line and the main line is $0.54\pm0.05$, whereas in the case of the 75\,nm thick Cr-capped \CFS\ film it is $0.41\pm0.05$. This finding suggests that this broad contribution is probably from the so called ``dead layer" near the substrate.\cite{Kallmayer2} It is well known that there is a substantial difference between lattice parameters of the MgO substrate (4.21\,\AA $\cdot \sqrt{2}$ = 5.95\,\AA) and \CFS\ (5.64\,\AA), which may yield a monotonous change in the structure of the Heusler film within a certain distance from the substrate. The factor $\sqrt{2}$ is attributed to the rotation of the \CFS\ unit cell by 45$^{\circ}$ with respect to the MgO unit cell.
The bulk of the film sample has an ideal $L2_1$ structure with a slight amount of Fe-Si off-stoichiometry, represented by the parameter $x$ in Co$_2$Fe$_{1+x}$Si$_{1-x}$. The values of $x$ in percent are listed in Table \ref{tab:samples}. All $x$ values were calculated by fitting a binomial distribution function to the
relative areas of the NMR resonance lines, as discussed in detail in Ref.\,\onlinecite{WKS2009}. Fig.\,\ref{fig:NMR} summarizes the relations between the different film characteristics (deposition temperature and saturation magnetization) and the results of the NMR analysis (linewidth, amount of off-stoichiometry and resonance frequency). For both types of capping layers, the amount of off-stoichiometry $x$ depends on the deposition temperature, specifically: the degree of off-stoichiometry decreases with increasing deposition temperature and hence with the degree of order (Fig.\,\ref{fig:NMR}(a,b)). Generally, the NMR linewidth is also a good measure of the structural order of a system. Fig.\,\ref{fig:NMR}\,c,d demonstrate a clear correlation between the magnetic moment of a given film and the NMR linewidth. Samples prepared at higher deposition temperatures generally have a larger moment and smaller NMR linewidth,
which is an indication of higher local order. In addition, there is a dependence of the resonance frequency of the main NMR line on the magnetic moment, samples with higher magnetic moment show a signal at lower frequencies, with the frequency closer to bulk \CFS\ (6\,$\mu_B$/f.u.). Such a trend is typically seen for NMR lines of Heusler compounds (compare Refs.\,\onlinecite{Wurmehl2012} and \onlinecite{Inomata2012}).

\subsubsection{XMCD of Si $L$-edge}
\label{subsec:XMCDSi}

\begin{figure}
\vspace*{0.0cm}
\includegraphics[width=0.95\columnwidth]{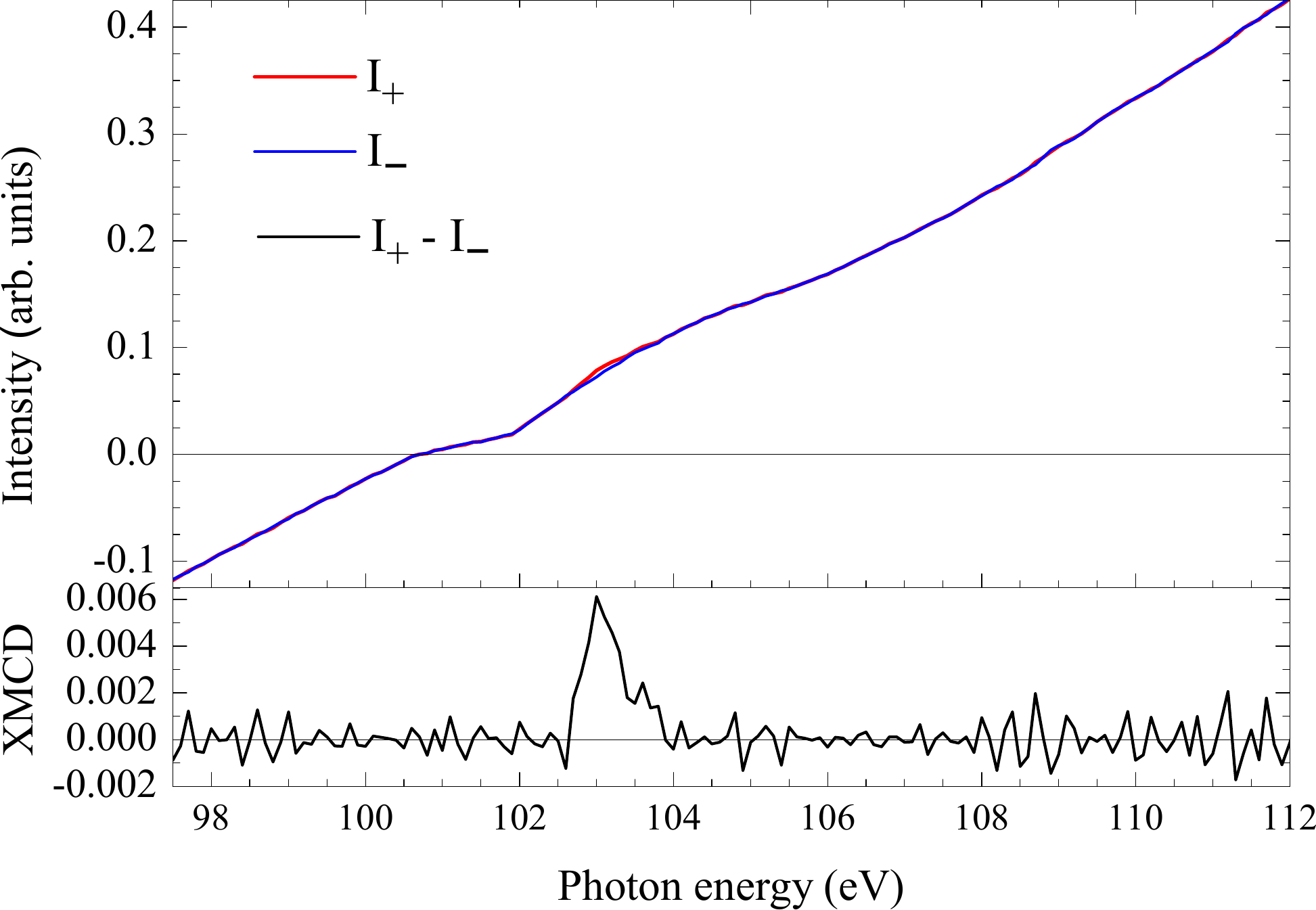} 
\hspace*{0cm}
\caption{\label{fig:SiXMCD} (Color online) X-ray absorption spectra and X-ray magnetic circular dichroism of \CFS (Cr4) measured at the Si $L_{3,2}$.}
\end{figure}

Although the total magnetization in \CFS\ is dominated by the large moments located at the Co and Fe atoms, it has been predicted \cite{Wurmehl} that a weak magnetic moment of ca. 0.13\,$\mu_B$ (LSDA+\textit{U}) is also induced at the Si atom with antiparallel orientation to the mean magnetization direction. The element-specific detection of this small magnetic moment remained a challenge because the Si $K$- and $L$-edges' signals are extremely weak. Recently, Antoniak \textit{et al}.\cite{Antoniak} showed the occurrence of an antiparallel Si moment of -0.01\,$\mu_B$ in thin Fe$_3$Si films using x-ray absorption at the Si $K$-edge and at the Si $L_{3,2}$ edges. Here, we investigated \CFS\ by using a similar technique. Fig.\,\ref{fig:SiXMCD} shows the XAS and XMCD data measured at the Si $L_{3,2}$ edges. We verified the result with field switching and polarization switching. In comparison to the data for Fe$_3$Si the XAS is much less structured revealing only a smooth increase near 102\,eV. The XMCD peak at 103\,eV shows a maximum value. The strength of the signal is given by the ratio of the difference of the intensities $I_{+}$, $I_{-}$ (0.006) and the value of the XAS spectra (0.1) at 103\,eV (values are extracted from Fig.\,\ref{fig:SiXMCD}). With $(I_{+}-I_{-})/\overline{I} = 0.006/0.1 = 0.06$, we calculate an asymmetry of 6\,\%. Calculation was performed according to Ref.\,\onlinecite{Antoniak}.
Notably, this value is about seven times larger than the value of 0.8\,\% obtained by Antoniak \textit{et al}.\cite{Antoniak}
As in the case of Fe$_3$Si, we conclude from the positive asymmetry that the orientation of the Si moment is antiparallel to the mean magnetization, which is in agreement with the prediction in Ref.\,\onlinecite{Wurmehl}.
Assuming that the size of the asymmetry scales with the magnetic moment we obtain an induced magnetic moment of -0.07\,$\mu_B$, which nicely agrees with our calculated value of Si total magnetic moment of -0.06\,$\mu_B$, ($\mu_{orb}=-0.0025 \mu_B$). As we probe Si {\it p-}states, the magnetization is induced to the hybridization of the magnetic 3d orbitals of the neighbors.
Since we could not measure at the $K$-edge we cannot conclude on the contribution from the orbital magnetic moment. Because the atomic structure in our case is cubic we may assume that the orbital moment contribution will be an order of magnitude smaller than the spin moment.

\subsection{Calculated results}

\subsubsection{XMLD at the $L_{3,2}$ edges of Fe and Co}

The calculation of adequate XMLD spectra at the Co and Fe $L_{3,2}$ edges of the full Heusler compound Co$_2$FeSi is still a challenging task. Using different exchange-correlation functionals, or employing additional parameters such as the Hubbard $U$, varying electronic structures can be obtained, leading to magnetic moments different from 6\,$\mu_B$ as well as to different x-ray spectra. 
Recently, the so-called fixed-spin method\cite{Wien2k} has been employed to constrain the magnetic moment of Co$_2$FeSi to the maximum measured value of 6\,$\mu_B$; the electronic structure was then computed selfconsistently under this constraint.\cite{Meinert} However, it seems that the low lying bands exhibit a non-physical exchange splitting of 1\,eV, see Fig.\,6 of Ref.\,\onlinecite{Meinert} (cf.\ Ref.\ \onlinecite{Bombor}). Another possibility to describe the high-spin electronic is to apply by the so-called Hubbard \textit{U} technique with the self-interaction correction\cite{Anisimov93} and choose suitable parameters, i.e.\ $U_{\rm eff}$(Co) = 4.8\,eV and $U_{\rm eff}$(Fe) = 4.5\,eV (see Ref.\,\onlinecite{Kallmayer}). 

 Recent electronic structure investigations have shown that the observed fine structure at the $L_3$ edges is often not exactly reproduced by {\it ab initio} calculations, whereas the $L_2$ edges often show just a simple oscillation and here the agreement with the first-principles calculations is good.\cite{Nolting10} The experimentally observed fine structure at at $L_3$ edge might stem from various, as yet unresolved origins. It might come from many-body effects or even be related to deficiencies of the sample.\cite{Nolting10}

\begin{figure}
\vspace*{0.0cm}
\includegraphics[width=0.95\columnwidth]{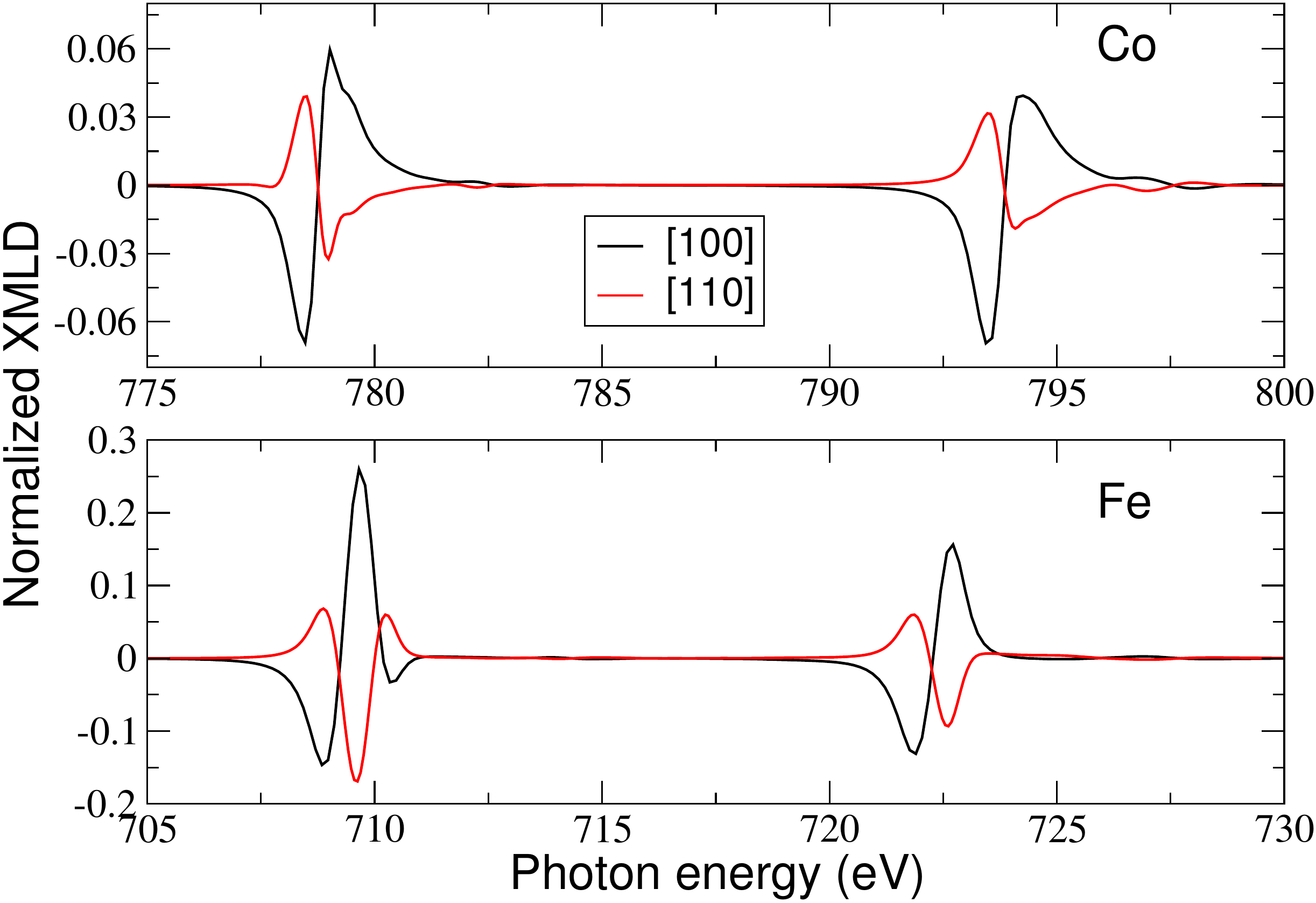}
\hspace*{0cm}
\caption{\label{fig:cfs-ab initio} (Color online) X-ray magnetic linear dichroism of \CFS\, calculated for the attenuation coefficient of the 75\,nm thick sample at the $L_{3,2}$ edge of Co (upper panel) and of Fe (lower panel), for the magnetization along the [100] or [110] directions. The calculations were performed with the LSDA+\textit{U} technique, using $U_{\rm eff}$ = 4.8\,eV(Co) and $U_{\rm eff}$ = 4.5\,eV(Fe). For details, see text.}
\end{figure}

As the LSDA+$U$ approach has given the best agreement between experimental and computed data, we employ here this approach. In Fig.\,\ref{fig:cfs-ab initio} we show the computed normalized XMLD spectra for both Co and Fe atoms and for the magnetization directions parallel to [100] or to [110]. The positions of the calculated $L_3$ edges have been aligned to the experimental $L_3$ edge energy.

Certain differences between the experimental and calculated spectra can be noted. For magnetization parallel to [100] the calculated XMLD spectrum at the Co $L_{3,2}$ edge does not exhibit a fine structure in contrast to the measured spectra shown in Fig.\,\ref{fig:overview100}. The calculations lack the drop (C) and the drop (D) on the high-energy side at the $L_3$ edge. For the $L_2$ edge the calculated XMLD spectrum exhibits a single oscillation with the same sign of the asymmetry of the two lobes as in the experiment, compare Fig.\,\ref{fig:cfs-ab initio} and Fig.\,\ref{fig:overview100}. For magnetization parallel to [110] the XMLD spectrum again exhibits a single oscillation at both the $L_3$ and $L_2$ edges as seen in the experiment having the correct asymmetry, \textit{i.e.} the positive peak is higher in magnitude than the negative peak. Moreover, the sign of the Co XMLD is inverted under the change of the magnetization direction from [100] to [110], in agreement with the previous study\cite{Kallmayer} and with studies of the pure $3d$ elements.\cite{Nolting10,Legut13} This inversion of the XMLD spectrum appears as a consequence of the crystal-field splitting of the $t_{2g}$ and $e_g$ states of the respective transition metal atom, with different states being sensitively probed for different magnetization directions.\cite{Kunes03} We note that the XMLD spectra of Co in Fig.\,4 of Ref.\,\onlinecite{Meinert} obtained with the LSDA+\textit{U} differ from ours by more fine structures at the $L_3$ and $L_2$ edges for both magnetization directions. However, it seems that we can reproduce such fine structures with calculations using the generalized gradient approximation with the Hubbard \textit{U} technique (GGA+\textit{U}) instead of LSDA+\textit{U}, but for the very similar value of $U_{\rm eff} = 4.8$\,eV for the Co atom.

The calculated Fe XMLD spectrum has a three peak structure at the $L_3$ edge, whereas a single oscillation at the $L_2$ edge exists for both magnetization directions, which again are inverted when changing the magnetization orientation from [100] to [110]. The experimental spectra at the $L_3$ edge exhibit similar features for magnetization along [100], except for peak D (see Fig.\,\ref{fig:overview100}\,e). The main disagreement between calculations and experimental data is in the magnitude of the first negative peak. It seems that the best agreement between calculated and measured XMLD spectra is achieved for the spectrum shown in Fig.\,\ref{fig:overview100}\,f. However, still the D feature (positive shoulder on the high-energy side) is not reproduced in the calculation. This was observed also in Ref.\,\onlinecite{Meinert}, even when various approximations for the exchange-correlation term were used. The single oscillation in the XMLD spectrum at the $L_2$ edge for magnetization along [100] is well reproduced.

A poorer agreement between measured and calculated XMLD spectra is found for both the Fe $L_{3,2}$ edges for magnetization along [110]. Experimentally, a single oscillation is observed here, while in ours and other theoretical calculations,\cite{Meinert} a three peak structure is found. Also, our calculated results differ from those of Kallmayer \textit{et al.}\cite{Kallmayer} For the $L_2$ edge again a single oscillation is computed, whereas experimentally predominantly a positive peak is detected, similar to Ref.\,\onlinecite{Meinert,Kallmayer}, and the negative peak is small. We ascribe this difference with Ref.\,\onlinecite{Kallmayer} to recent improvement of our calculations of the dielectric tensor elements (complex linear response formula) and the present use of a more sophisticated multilayer model for describing the light-sample interaction (Fresnel theory using Vi\v{s}\v{n}ovsk\'{y}'s formalism\cite{Vis91}) over the simpler approach used in Ref.\,\onlinecite{Kallmayer}. 

The calculated XMLD spectra are about one order of magnitude larger than the experiment. This is understandable in the TEY experiment, as we do not account for any loss due to scattering within the sample. We have tested the effect of substrate (150\,nm MgO) as well as 3\,nm capping layer of Al or Cr, where in addition to the {\it ab initio} calculated Co$_2$FeSi optical data, we adopted for the capping and substrate experimental data from the x-ray database.\cite{Henke} We found that none of those influence the spectra in the energy range of the Fe and Co $L$-edges.

\subsubsection{XMCD at the $L_{3,2}$ edges of Si}

\begin{figure}
 \vspace*{0.0cm}
 \includegraphics[width=0.95\columnwidth]{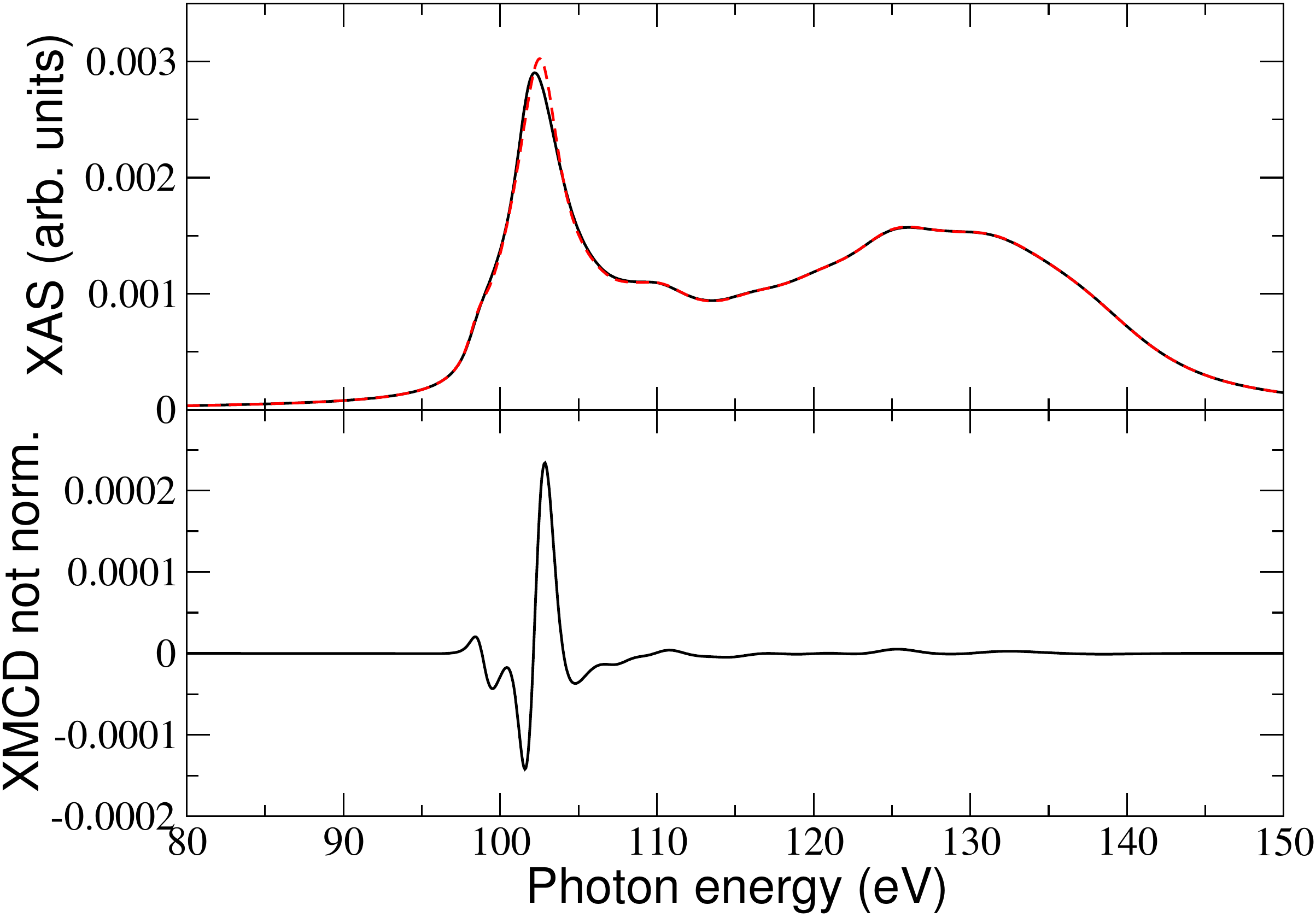} 
 \hspace*{0cm}
 \caption{\label{XMCD-Si} (Color online) Calculated X-ray absorption spectra and X-ray magnetic circular dichroism (not normalized) at the $L_{3,2}$ edges of Si in \CFS. See text for details.}
 \end{figure}

Using the LSDA+\textit{U} approach for the electronic structure of \CFS, we also calculated the XAS and XMCD at the $L_{3,2}$ edges of Si. The results are given in Fig.\,\ref{XMCD-Si}. The XMCD strength is calculated analog to section \ref{subsec:XMCDSi}. From Fig.\,\ref{XMCD-Si} at 103\,eV we obtain $0.0002/0.003 = 0.1$. It means that XMCD is about 10\,\% of the XAS magnitude, which is in agreement with the experimental finding of 6\,\%. Note that experimental XMCD spectrum appears to exhibit a stronger broadening than the computed spectrum.
The computed spectra only shows contribution from Si, whereas experimentally we see contribution from Co and Fe, which create a huge background signal due to the weak Si signal. We note that it is a general problem to subtract the background reliably. Consistent with the experiment, we find that an $L_2$ edge XMCD signal is not present. The calculated XMCD spectrum is overall similar in shape to that calculated for Si in Fe$_3$Si by first-principles methods in Ref.\,\onlinecite{Antoniak}, except that here we do not obtain a negative peak at the high-energy side.

\section{Summary and Conclusion}
In summary, magnetic properties of two series of \CFS\ films, one capped with Cr and the other with an Al layer, were investigated. Both series comprise four films, each exhibiting a different local order and varied saturation magnetization. A comparison of the recorded XMLD spectra revealed that the Al-capped films show large variations, whereas the Cr-capped series exhibit consistent features within the series. The XMLD spectra of a \CFSA\ film showed similar features like Al-capped \CFS. Thus, an intermixing of Al atoms of the \CFS\ subsurface during deposition was assumed. This is a finding which needs to be considered carefully when making tunnel barriers for TMR devices, since it is important to have a stoichometric structure at the interface. For a systematic investigation, certain points were taken as a scale for the XMLD strength and the values were plotted versus the saturation magnetization of the films. An expected rise was noted for the Cr-capped \CFS\ series, whereas the Al-capped series showed no dependence.
All samples show a clear dependence of the film quality on the deposition conditions, while the deposition temperature has the
highest impact on the film quality. By increasing the deposition temperature the
films are better ordered and furthermore the off-stoichiometry is reduced.

Comparison with \textit{ab initio} calculated XMLD spectra showed an overall good agreement, however, certain differences remain. Whether these are due to the residual off-stoichiometry of the films or if further improvement of the calculation is needed is currently an open question. Experimentally, the feature-rich spectra of the Cr-capped samples change only qualitatively with disorder. This observation still needs to be reproduced in future calculations.
The magnetic moment at the Si atom in \CFS\ could be detected with XMCD. A small magnetic contribution aligned antiparallel to the Co moments was observed, in agreement with \textit{ab initio} calculations.

\begin{acknowledgments}
We acknowledge support by DFG research group ASPIMATT DFG Ja821/5-1 and Graduate School of Excellence Materials Science in Mainz (MAINZ) - GSC266.
We thank S. Cramm for support at BESSY. SW gratefully acknowledges funding by Deutsche Forschungsgemeinschaft DFG under the Emmy Noether programme in project WU595/3-1. DL acknowledges support within projects Reg. No. CZ.1.07/2.3.00/20.0074 and Reg. No. CZ.1.05/1.1.00/02.0070, both supported by Operational Programme 'Education for competitiveness' funded by Structural Funds of the European Union and state budget of the Czech Republic. AK acknowledges DFG (Graduate School of Excellence Materials Science in Mainz (MAINZ) - GSC266).
\end{acknowledgments}


\end{document}